# Sub-Diffractive Band-Edge Solitons in Bose-Einstein Condensates in Periodic Potentials

Kestutis Staliunas[1], Ramon Herrero[2] and Germán J. de Valcárcel[3]

[1]*Institució Catalana de Recerca i Estudis Avançats (ICREA), Departament de Física i Enginyeria Nuclear, Universitat Politècnica de Catalunya, Colom 11, 08222 Terrassa, Spain*

[2]*Departament de Física i Enginyeria Nuclear, Universitat Politècnica de Catalunya, Comte d'Urgell 187, 08036 Barcelona, Spain*

[3]*Departament d'Òptica, Universitat de València, Dr. Moliner 50, 46100 Burjassot, Spain*

A new type of matter wave diffraction management is presented that leads to sub-diffractive soliton-like structures. The proposed management technique uses two counter-moving, identical periodic potentials (e.g. optical lattices). For suitable lattice parameters a novel type of atomic band-gap structure appears in which the effective atomic mass becomes infinite at the lowest edge of an energy band. This way normal matter-wave diffraction (proportional to the square of the atomic momentum) is replaced by fourth-order diffraction, and hence the evolution of the system becomes sub-diffractive.

Dispersion management for atomic matter waves in periodic potentials is an effervescent research area in which rich nonlinear wave phenomena, such as the existence of gap solitons in the atomic band-gap structure, have been predicted [1] and observed [2]. In these studies the main focus is laid on the vicinity of the upper band-edge, where the effective atomic mass is negative, and enables, e.g., the excitation of localized wave packets of Bose-Einstein condensates (BECs) with repulsive interaction, as demonstrated theoretically [3] and experimentally [4].

Less attention has been paid so far to the bulk areas of the propagation band, especially to the vicinity of the inflection point of the dispersion curve ($\partial^2 E/\partial k^2 = 0$, where $E(k)$ represents the dispersion relation between the energy $E$ and quasimomentum $k$ of the condensate). At this point the effective mass $m^* = \hbar^2/(\partial^2 E/\partial k^2)$ [5] of the condensed particles becomes infinite or, in other words, "normal" diffraction is suppressed. We call this special point a "zero-diffraction point" (ZDP). This ZDP occurs at a special value of the relative quasimomentum $k = k_{ZDP}$ between the BEC and the periodic lattice, which depends on the strength of the potential. Regimes of opposite (positive or negative) diffraction occur for quasimomenta at opposite sides of $k_{ZDP}$. The ZDPs so far analyzed in linear [6], nonlinear [7], and dissipative [8] cases are asymmetric, in the sense that odd-order derivatives of the dispersion relation are nonzero at $k = k_{ZDP}$. This means that, apart from a drift, the dominating role is played by the third-order dispersion term and this seems to exclude the possibility of stable solitons in the vicinity of the ZDP [9].

The fact that a ZDP appears at $k = k_{ZDP}$ means that, if the lattice is at rest in the laboratory frame, the BEC must be given a velocity $v = \pm v_{ZDP}$, $v_{ZDP} = \hbar k_{ZDP}/m$, in order to be in a ZDP (both signs are allowed due to the reflection symmetry of the problem). However, owed to the Galilean invariance of the problem, the same situation is reached if the BEC is at rest in the laboratory frame ($k = 0$) and the lattice moves with velocity $\pm v_{ZDP}$. This reasoning allows envisaging a way to restore the parity symmetry of the system around the ZDP, consisting in the use of two counter-moving periodic lattices, which lead to the following form for the potential:

$$V(x,t) = V_0[\cos(k_0 x - \omega_0 t) + \cos(k_0 x + \omega_0 t)]. \quad (1)$$

Intuitively we can expect that if the velocity of the lattices ($v = \omega_0/k_0$) is close to $v_{ZDP}$, a resting BEC wavepacket ($k = 0$) will be in a ZDP as in this case the BEC is in a ZDP for both lattices separately. Thus one can expect that $k_{ZDP} = 0$, in which case the dominant term of the dispersion relation will be the fourth-order one, owed to the reflection symmetry of the problem. In the rest of this Letter we shall prove numerically and analytically this conjecture. We shall analyze the consequences of this "subdiffractiveness" and demonstrate, analytically and numerically, the existence of stable "subdiffractive" matter-wave soliton-like solutions.

A one-dimensional (1D) BEC is considered, corresponding to a cigar-shape condensate with a strong trapping potential in the radial direction [10]. The starting point of our analysis is the effective 1D Gross-Pitaevskii equation for such BEC subjected to potential (1) in the longitudinal direction $x$ and no other trapping potential in that direction [11],

$$i\hbar\partial_t\psi = \left[-\hbar^2/(2m)\partial_x^2 + V(x,t) + g|\psi|^2\right]\psi, \quad (2)$$

where $g = \hbar \nu k_0 a$ is the effective two-body interaction coefficient, being $\nu$ the angular frequency of the radial trap and $a$ the interatomic $s$-wave scattering length ($a > 0$ for a repulsive BEC, which we consider). Equation (2) is complemented with the normalization condition $k_0 \int dx |\psi(x,t)|^2 = N$, $N$ is the number of particles, which renders $\psi(x,t)$ adimensional. The study is facilitated by adopting the following dimensionless quantities,

$$X = k_0 x, T = \varpi t, f = \frac{V_0}{2\hbar\varpi}, \Omega_0 = \frac{\omega_0}{\varpi}, G = \frac{g}{\hbar\varpi}, \quad (3)$$

where $\varpi = \hbar k_0^2/(2m)$, in terms of which Eq. (2) becomes

$$i\partial_T\psi = \left[-\partial_X^2 + 4f\cos X\cos(\Omega_0 T) + G|\psi|^2\right]\psi, \quad (4)$$

with normalization $\int dX|\psi(X,T)|^2 = N$.

First we study the linear dispersion properties of Eq. (4) by expanding $\psi(X,T)$ as [12]:

$$\psi(X,T) = e^{i(KX-\Omega T)} \sum_m \sum_n \psi_{m,n} e^{i(mX+n\Omega_0 T)}. \quad (5)$$

In physical units $p = \hbar k = \hbar k_0 K$ and $E = \hbar\omega = \hbar\varpi\Omega$ are the quasimomentum and the quasienergy of the BEC, respectively. Substitution of (5) into (4) with $G = 0$ (linear response) yields the following system of equations:

$$[(m+K)^2 + n\Omega_0 - \Omega]\psi_{m,n} - f \sum_{p=m\pm 1}\sum_{q=n\pm 1}\psi_{p,q} = 0. \quad (6)$$

Solvability condition of (6) results in a dispersion relation $\Omega(K)$ of which Fig. 1 is an example. In the limit of vanishing periodic potential ($f = 0$) the solution to (6) consists of the family of parabolas $\Omega(K) = (m+K)^2 + n\Omega_0$, $m,n \in Z$ (dashed curves in Fig. 1) shifted one with respect to another by the reciprocal vectors of the lattice of the space-time periodic potential that, with the used normalizations, read $(\pm 1, \pm \Omega_0)$. The presence of the potential ($f \neq 0$) lifts the degeneracy at the crossing points and gives rise to band-gaps as usual. What is most relevant to our study is that for a special relation between the potential parameters $(f, \Omega_0)$ a plateau around $K = 0$ appears in one of the bands, as shown in Fig. 1. This plateau corresponds to $\partial^2\Omega/\partial K^2 = 0$, which thus signals an infinite effective mass and, accordingly, the vanishing of diffraction (to the leading order). Thus one of the bands contains a ZDP at $K = 0$ as conjectured and this opens the way to new phenomena in matter wave dynamics.

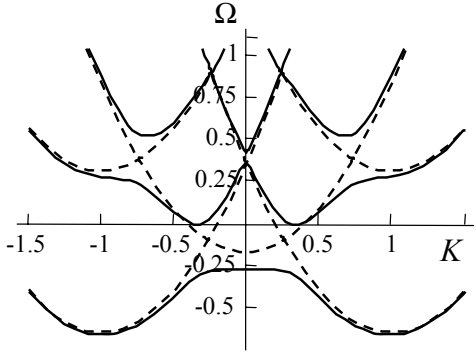

**FIG. 1.** Dimensionless quasienergy $\Omega$ as a function of dimensionless quasimomentum $K$ for a normalized lattices' velocity $\Omega_0 = 0.48$. Dashed lines: Limit of vanishing periodic potential ($f = 0$). Solid lines: $f = 0.215$. The lowest solid line corresponds to the band containing a zero diffraction point (ZDP) at $K = 0$, which manifests as a plateau of the dispersion curve. Also neighboring bands are shown.

In order to gain insight into the characteristics of the ZDP in some limit we consider the case of weak potential $f \to 0$. Our numerical analysis of Eq. (6) in that case indicates that a ZDP appears if $\Omega_0 \to 1$. Moreover, only amplitudes $\psi_{0,0}$, $\psi_{1,-1}$, and $\psi_{-1,-1}$ in (5) attain appreciable values. Guided by this result we use then a truncated version of (5) restricted to those amplitudes that, obviously, only captures the behavior of the band containing the ZDP. The solvability of the corresponding Eq. (6) can be cast now as

$$\Omega = K^2 + f^2 W(K,\Omega), \quad (7)$$

where $W(K,\Omega) = \frac{2w}{w^2 + 4K^2}$ and $w = \Omega - K^2 + \Omega_0 - 1$. Our goal is to find which relation between the potential parameters, $f$ and $\Omega_0$, must be fulfilled in order that $\partial^2\Omega/\partial K^2 = 0$ at $K = 0$. As we are considering the limit $f \to 0$, Eq. (7) can be solved approximately as $\Omega = K^2 + f^2 W(K,K^2)$. For small $K$ one has,

$$\Omega = D_0 + D_2 K^2 + D_4 K^4 + O(K^6), \quad (8)$$

where $D_0 = \frac{2f^2}{\Omega_0 - 1}$, $D_2 = 1 + \frac{8f^2}{(\Omega_0 - 1)^3}$ and $D_4 = \frac{32f^2}{(\Omega_0 - 1)^5}$. The condition for vanishing diffraction $D_2 = 0$ leads to

$$8f^2 = (1 - \Omega_0)^3, \quad (9)$$

which, as we have assumed $f$ small, forces $\Omega_0$ to be close to unity, in agreement with the numerics. We call Eq. (9) the "zero diffraction curve" (ZDC) for obvious reasons. Figure 2 represents the ZDC as obtained by the previous analysis (dashed line), and as obtained by solving numerically Eq. (6) truncated to 5 mode amplitudes (solid line). Equation (9) reproduces the correct behavior of the true ZDC for small $f$, as expected. According to Fig. 2 there exists a minimum value of the normalized potential velocity $\Omega_0$ below which no tuning of the potential strength $f$ can lead to the formation of a ZDP. Also note that the form of the band changes qualitatively along the ZDC around $f \approx 0.6$ according to the insets, where a point where even the fourth order diffraction disappears seems to exist. These two features are not captured by our asymptotic analysis (7)-(9), and are not to be considered in the present Letter.

Just at the ZDC, diffraction vanishes at $K = 0$ and the (linear) BEC wavefunction (5) corresponding to the band (8) takes a special form, which can be computed from Eq. (6), see below. This solution is the Bloch mode associated with the ZDP. A slight departure of the parameters from the ZDC as well as the presence of the nonlinearity in Eq. (4) (if weak) will give rise in general to a solution that can be seen as a slow (in space and in time) modulation of the nondiffractive Bloch mode.

As a second step we wish to determine the evolution equation for that slow modulation. This can be done in the considered limit, $\Omega_0 = 1 - \delta$, $|\delta| \ll 1$, in the vicinity of the ZDC, $8f^2 = \delta^3 + \mu\delta^4$, where $\mu$ controls the distance from the ZDC (9). The derivation follows a standard multiscale asymptotic expansion of Eq. (4) that uses $\delta$ as the smallness parameter. Proper scalings for space and time can be set by analyzing the behavior of the dispersion curve (8) in this limit, which yields $D_0 \propto \delta^2$, $D_2 \propto \delta$ and $D_4 \propto \delta^{-2}$. As we wish to construct an equation that captures the dynamics of the BEC in a consistent way even when $D_2 = 0$, the sought equation must contain at least terms proportional to the fourth-order spatial derivative (in correspondence with the term $D_4 K^4$ in the dispersion relation). Thus we must impose that $D_2 K^2$ and $D_4 K^4$ are of the same order in $\delta$, which yields $K = O(\delta^{1/2})$. This scaling indicates that the envelope of the Bloch mode should vary on a spatial scale on the order of $\delta^{-3/2}$. Introducing all these scalings into Eq. (7) and solving it in powers of $\delta$ yields

$$\Omega = \overline{\Omega} - d_2 K^2 - d_4 K^4 + O(\delta^5), \qquad (10)$$

where $4\overline{\Omega} = \delta^2 + (1/4 - \mu)\delta^3(1 - \delta/2)$, $d_2 = (\mu - 1)\delta$ and $d_4 = 4/\delta^2$.

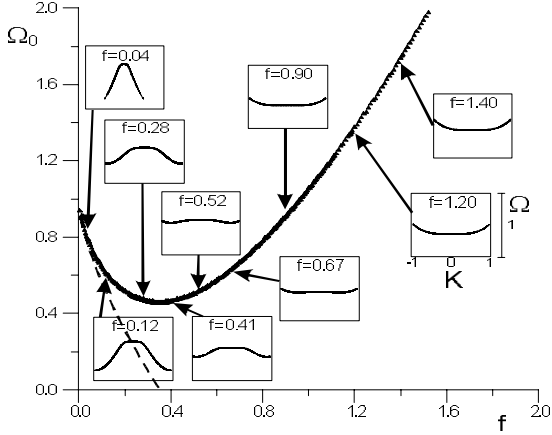

**Fig. 2**. Zero diffraction curve relating the values of the normalized lattices velocity $\Omega_0$ with the normalized potential strength $f$ at which a ZDP appears at $K = 0$, as obtained by numerically solving Eq. (6). Insets display the first branch of the dispersion diagram around $K = 0$. The dashed line corresponds to Eq. (9), valid at small values of $f$.

Equation (10) agrees, to the leading order, with Eq. (8) and represents a better approximation as now a single Taylor expansion (in terms of $\delta$) has been done that affects simultaneously all parameters $(\Omega_0, f, K)$. We see that the scalings are indeed consistent as they are able to contain nontrivial information about the quasimomentum (proportional to $K$), even at the ZDC that now reads $\mu = 1$, i.e. $8f^2 = \delta^3(1 + \delta)$, which is the first order correction to the ZDC (9). The last information we need is about the form of the Bloch mode. Application of previous scalings to Eq. (6) truncated to the three used amplitudes yields $\psi_{1,-1} = \psi_{-1,-1} = -\sqrt{\delta/8}\,\psi_{0,0}$, to the leading order, which sets the form of the linear Bloch mode at the ZDP as $B(X,T) = 1 - \sqrt{\delta/2}\,e^{-i\Omega_0 T}\cos X$. Now we can search a solution to Eq. (2) in the form

$$\psi(X,T) = \Psi(X,T)B(X,T)e^{-i\overline{\Omega}T} + O(\delta^{5/4}) \qquad (11)$$

The exponential is introduced in order to get rid of the trivial time evolution, that not depending on $K$ in Eq. (10), and $\Psi(X,T) = \delta^{3/4}\Phi(\delta^{1/2}X, \delta^4 T)$, with $\Phi$ a slowly varying function of order $\delta^0$, is the envelope whose evolution equation is to be sought. The scale for $\psi$ is a consequence of the normalization condition. It reads $\int dX|\Psi(X,T)|^2 = N$ to the leading order. The fact that we are interested in localized solutions, whose width $\Delta X \propto \delta^{-3/2}$, determines finally the scaling. Finally we impose that the nonlinear term be of order $\delta^4$ in order to be able to compensate dispersion, what implies $G \propto \delta^{5/2}$. Substitution of all previous scalings into Eq. (4) leads to a hierarchy of equations at increasing powers in $\delta^{1/2}$. Going up to the fifth order one finds

$$i\partial_T \Psi = d_2 \partial_X^2 \Psi - d_4 \partial_X^4 \Psi + G|\Psi|^2 \Psi, \qquad (12)$$

where $d_2$ and $d_4$ are as in (10).

Equation (12) is the central result of our analysis as it describes the character of the BEC close to the ZDC. Note that at $\mu = 1$ the normal diffraction term disappears, in agreement with the analysis of (10). Then the effective "kinetic energy" term becomes proportional to the fourth power of the quasimomentum. Also note that the nonlinear term in (12) is unchanged with respect to Eq. (4), unlike in [13]. The reason is that in [13] the considered Bloch mode had a strong spatial modulation while in our case $B(X,T) = 1 + O(\delta^{1/2})$, $\delta \ll 1$. Far from this limit the coefficient of the nonlinear term should be smaller than $G$ in a nontrivial way.

A remarkable property of Eq. (12) is that it supports bell-shaped solitons, some of which have analytic form [15]. The validity of (12) is to be checked by comparing its results with those obtained by integrating Eq. (4). Most interesting consequences of (12) follow from the scaling characteristics of its solutions:

(i) Far away from the ZDC, when the second derivative dominates, the usual Gross-Pitaevskii equation is recovered that sustains exact sech (tanh) solitons for $d_2 G > 0$ ($d_2 G < 0$). If we denote by $\Delta X$ the width of the soliton, and by $|\Psi_0|^2$ the peak value of the atom density, the family of solutions is then given by $d_2/\Delta X^2 \propto |\Psi_0|^2$. On the other hand, as $N = \int dX|\Psi(X,T)|^2 \propto \Delta X|\Psi_0|^2$ for any bell shaped wavefunction, we obtain $N\Delta X \propto d_2$. Hence this case is characterized by the power-law

$$\log_{10} \Delta X = a_2 - \log_{10} N, \qquad (13)$$

$a_2 \propto \log_{10} d_2$. (ii) On the ZDC $d_2 = 0$, and the solitonic family verifies $d_4/\Delta X^4 \propto |\Psi_0|^2$ or, alternatively, $N\Delta X^3 \propto d_4$. The power-law is hence now

$$\log_{10} \Delta X = a_4 - \tfrac{1}{3}\log_{10} N, \qquad (14)$$

$a_4 \propto \log_{10} d_4$. (iii) Close to the ZDC both normal and fourth-order diffraction play a role and one must expect $d_2/\Delta X + d_4/\Delta X^3 \propto N$. For small (large) $N$ the dominant term is the one with $d_2$ ($d_4$); hence we expect a smooth transition from (13) to (14) by increasing $N$.

We checked the above predictions by numerical integration of Eq. (4). Figure 3 shows the relation between $\Delta X$ and $N$ for four families of solitons corresponding to four different values of $\Omega_0$. We find two different linear regimes in the log-log plot indicating the existence of two different power-law scalings as predicted.

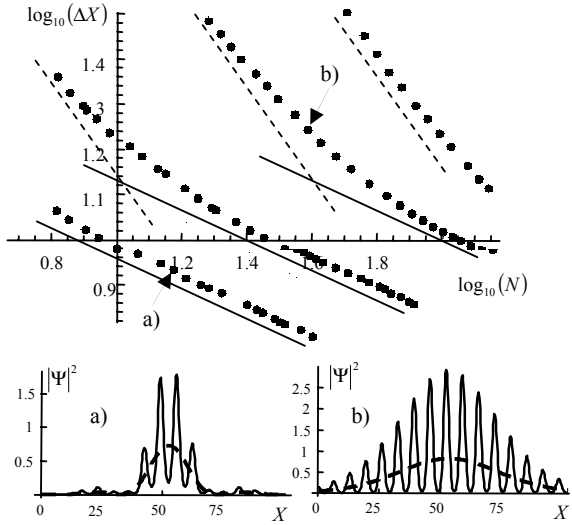

**FIG. 3**. Top: Relation between the width $\Delta X$ and the number of atoms $N$ of solitons (log-log scale) for $f = 0.1$ and $G = 0.004$. The four groups of points correspond, from top-right to low-left, to four different lattices velocities: $\Omega_0 = 0.89$, 0.79, 0.69, and 0.59, respectively. Dashed (solid) lines correspond to power-law scalings of normal (nondiffractive) solitons. Bottom: Soliton spatial profiles corresponding to points a) and b) in the top figure; dashed lines show the envelope of the Bloch mode, obtained by Fourier filtering. The ZDP, as follows from the linear theory (6) occurs for $\Omega_0 = 0.595$ (see Fig. 2).

The corresponding slopes far from the ZDC (right-top of the figure) follow the scaling of normal solitons, Eq. (13), while very close to the ZDC (the left-bottom of the figure) the corresponding slopes follow the sub-diffractive scaling, Eq. (14). It is remarkable that along any solitonic family, the subdiffractive power-law (14) applies for sufficiently large $N$, as anticipated, meaning that nonlinearity favors the existence of subdiffractive solitons. This evidences that Eq. (12), even if derived for weak nonlinearity and for parameters close to the ZDC, is able to qualitatively capture the behavior of the BEC under situations far from the considered asymptotic limit.

Subdiffractive solitons (residing exactly on the ZDC), or mixed ones (above the ZDC) are stable according to our numerical analysis. We observe that if we inject a "condensate droplet" of given width $\Delta X$, whose number of atoms $N$ is larger than the corresponding solitonic value (that shown in Fig. 3), then the droplet gets rid of the excess atoms by waves of continuous radiation until a solitonic state is reached. If $N$ exceeds more than twice the solitonic value, then the splitting into two solitons is observed. We note that we were unable to excite stable solitons below the ZDC, where $d_2 < 0$ and that ensembles of repulsing atoms ($G > 0$) do not result is stable bright solitons in this case, as expected.

Concluding, we have demonstrated that the use of counter-moving periodic potentials is a powerful tool for the controllable diffraction management of BECs. In particular, the proposed scheme can eliminate second order (usual) diffraction, which is then replaced by fourth order diffraction, leading to stable subdiffractive solitons. The suggested scheme is applicable to two- and three-dimensional condensates (correspondingly two- and three pairs of the counter-moving lattices are to be used). The analysis, similar to that performed here, results in stable solitons (absence of collapse) both in 2 and 3 spatial dimensions. Finally we note that the diffraction management technique put forward in this Letter can be applied to the light propagation through photonic crystals with Kerr-like nonlinearities.

This work was financially supported by the Spanish Ministerio de Educación y Ciencia and European Union FEDER through projects FIS2004-02587 and BFM2002-04369-C04-01.